\documentclass[aps, prl, twocolumn,showkeys,showpacs,amsmath,amssymb,superscriptaddress]{revtex4}
\usepackage{graphicx} 
\usepackage{dcolumn} 
\usepackage{bm}
\usepackage{epsfig}
\usepackage{subfigure}

\begin{document}
\title{Pre-Turbulent Regimes in Graphene Flows}

\author{M. Mendoza} \email{mmendoza@ethz.ch} \affiliation{ ETH
  Z\"urich, Computational Physics for Engineering Materials, Institute
  for Building Materials, Schafmattstrasse 6, HIF, CH-8093 Z\"urich
  (Switzerland)}

\author{H. J. Herrmann}\email{hjherrmann@ethz.ch} \affiliation{ ETH
  Z\"urich, Computational Physics for Engineering Materials, Institute
  for Building Materials, Schafmattstrasse 6, HIF, CH-8093 Z\"urich
  (Switzerland)}

\author{S. Succi} \email{succi@iac.cnr.it} \affiliation{Istituto
  per le Applicazioni del Calcolo C.N.R., Via dei Taurini, 19 00185,
  Rome (Italy),\\and Freiburg Institute for Advanced Studies,
  Albertstrasse, 19, D-79104, Freiburg, Germany}

\date{\today}
\begin{abstract}
  We provide numerical evidence that electronic pre-turbulent
  phenomena in graphene could be observed, under current experimental
  conditions, through detectable current fluctuations, echoing the
  detachment of vortices past localized micron-sized impurities.
  Vortex generation, due to micron-sized constriction, is also
  explored with special focus on the effects of relativistic
  corrections to the normal Navier-Stokes equations. These corrections
  are found to cause a delay in the stability breakout of the fluid as
  well as a small shift in the vortex shedding frequency. Finally, a
  relation between the Strouhal number, a dimensionless measure of the
  vortex shedding frequency, and the Reynolds number is provided under
  conditions of interest for future experiments.
\end{abstract}



\maketitle

Since its recent discovery \cite{natletter,Geim1}, graphene has
continued to surprise scientists with an amazing series of spectacular
properties, such as ultra-high electrical conductivity, ultra-low
viscosity to entropy ratio, combination of exceptional structural
strength and mechanical flexibility, and optical transparency. Many of
these fascinating effects are due to the fact that, consisting of
literally a single carbon monolayer, graphene represents the first
instance of a truly two-dimensional material (the ``ultimate
flatland'' \cite{PhysToday}). Moreover, due to the special symmetries
of the honeycomb lattice, electrons in graphene are shown to behave
like an effective Dirac fluid of {\it massless} chiral quasi-particles
propagating at a Fermi speed of about $v_F \sim c/300 \sim 10^6$
m/s. This configures graphene as a very special, slow-relativistic
electronic fluid, where many unexpected quantum-electrodynamic
phenomena can take place, \cite{QGP-1,QGP-2,QGP-3}.
%such as zero-reflectivity barrier penetration (Klein paradox), which would
%otherwise be impossible in standard quantum electro-dynamics.
%Spectacular effects are not confined to the quantum realm. 
% For instance, since electrons are about $300$ times slower than the
%photons they exchange, their mutual interaction is proportionately
%enhanced, corresponding to an effective fine-structure constant
%$\alpha_{gr} = e^2/\hbar v_F \sim 1$.  As a result of such strong
%interactions, it has been recently proposed that this peculiar 2D
%graphene electron gas, should be characterized by an exceptionally low
%viscosity/entropy ratio (near-perfect fluid), coming close to the
%famous AdS-CFT lower bound conjectured for quantum-chromodynamic
%fluids, such as quark-gluon plasmas \cite{QGP-1,QGP-2,QGP-3}. 
In particular, the capability of reaching down viscosity to entropy
ratios smaller than that of superfluid Helium at the lambda-point, has
recently spawned the suggestion that electronic transport in graphene
may support pre-turbulent phenomena, Ref. \cite{grapPRL}.

\begin{figure}  
  \centering
  \subfigure[]{
    \includegraphics[scale=0.15]{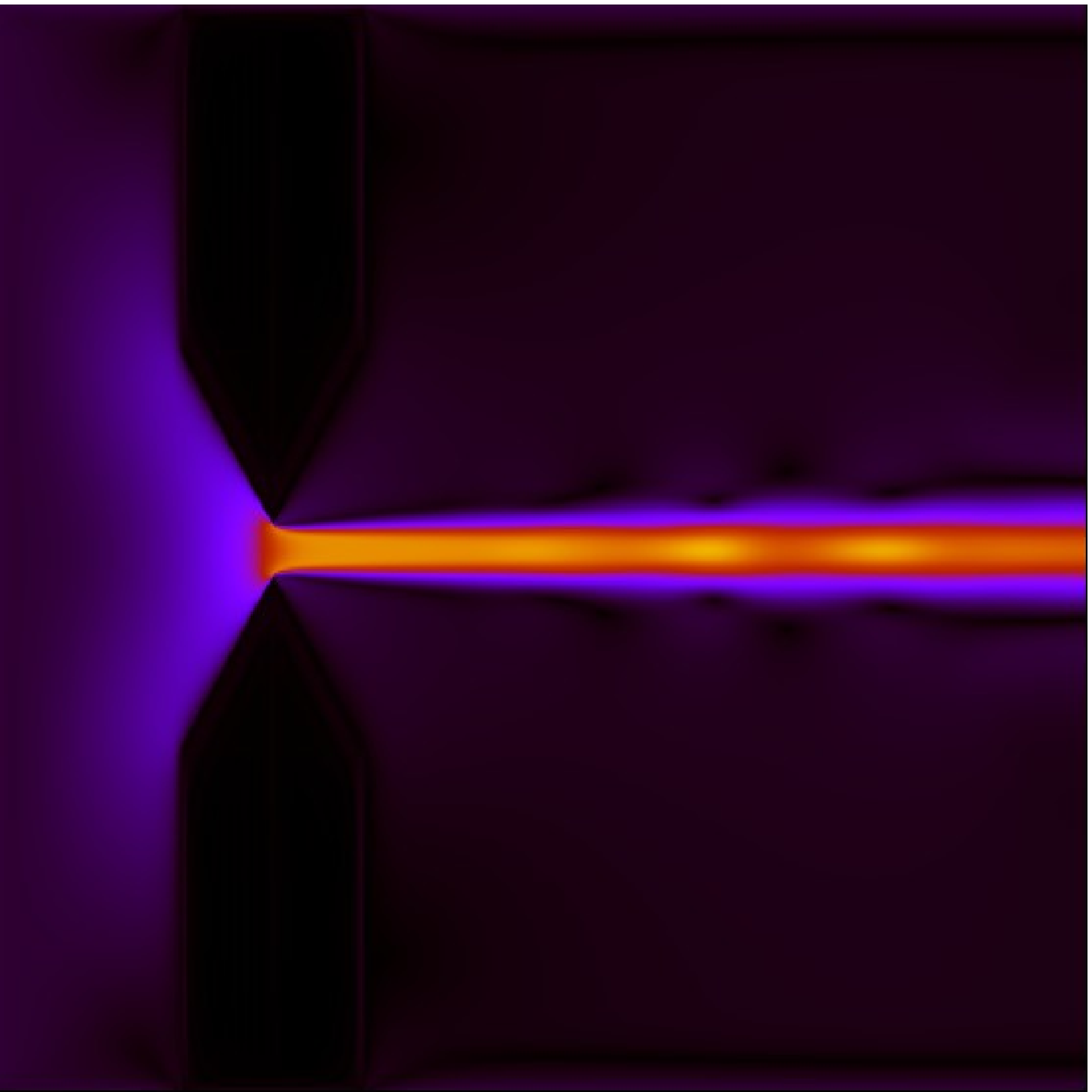}
    \label{plot5a}
  }\subfigure[]{
    \includegraphics[scale=0.15]{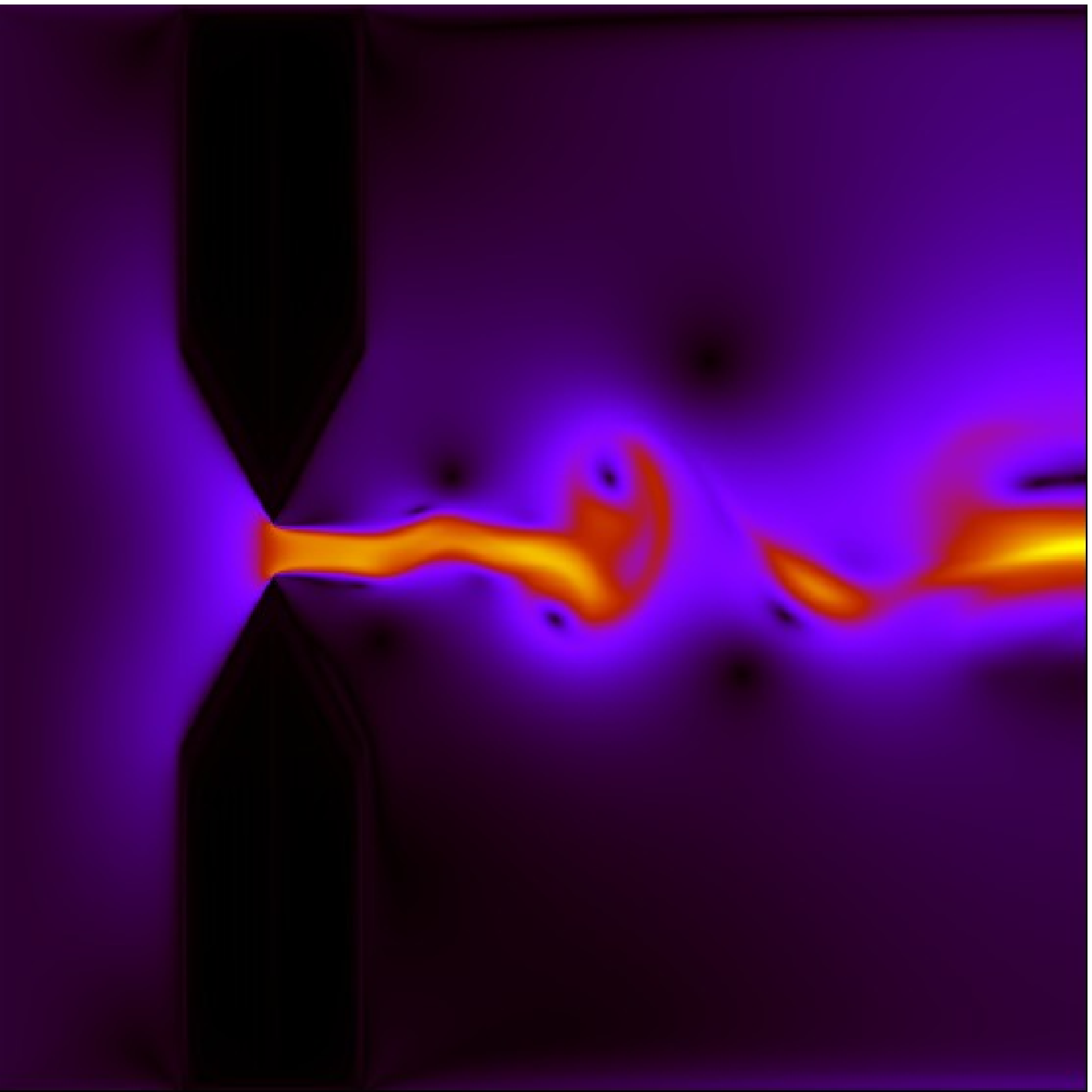}
    \label{plot5b}
  } 
  \subfigure[]{
    \includegraphics[scale=0.15]{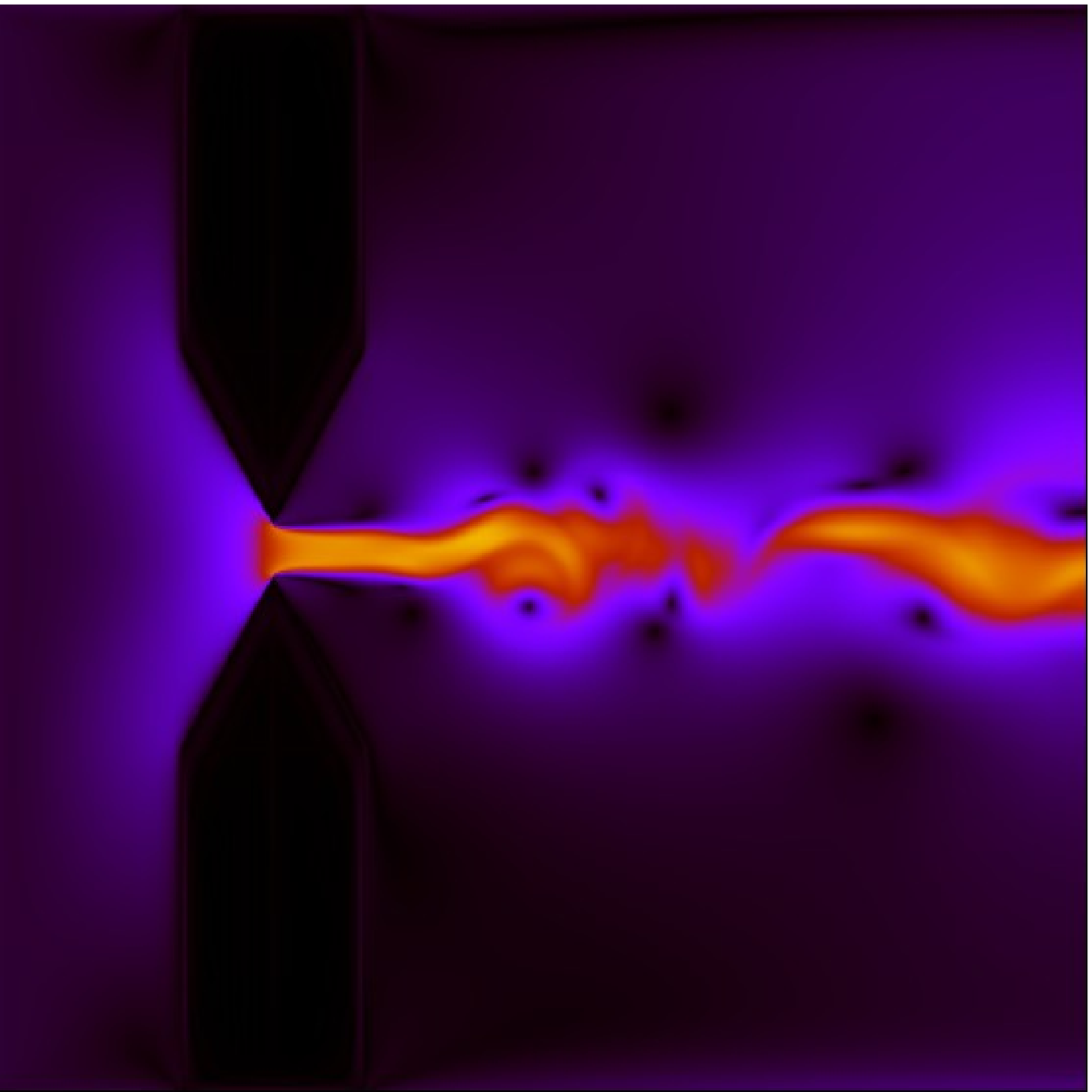}
    \label{plot5c}
  }\subfigure[]{
    \includegraphics[scale=0.15]{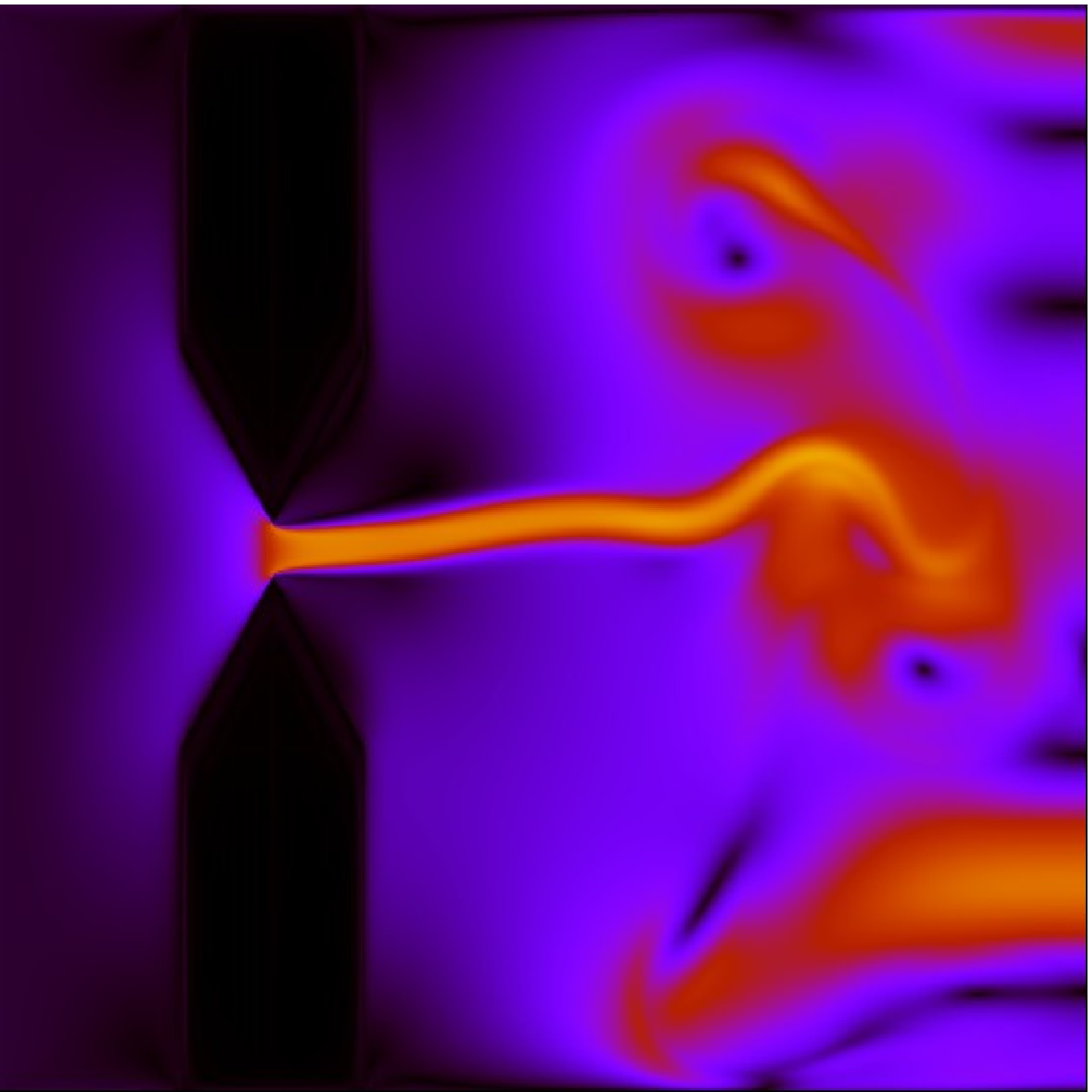}
    \label{plot5d}
  }\caption{Pre-turbulence at Reynolds number $Re=25$ in graphene is
    shown: at $379400$ time steps, \subref{plot5a} and
    \subref{plot5b}; and at $603400$ time steps, \subref{plot5c} and
    \subref{plot5d}. For \subref{plot5b} and \subref{plot5d} the term
    $\partial p /\partial t$ was removed. The color represents the
    magnitude of the velocity.}\label{plot5}
\end{figure}
In this Letter, we pursue this suggestion in quantitative terms.  More
precisely, we simulate the relativistic graphene-fluid equations,
proposed in Ref. \cite{grapPRL}, under conditions of present and
prospective experimental realizability. Our main result is that
micro-scale impurities, as small as a few microns, are capable of
triggering coherent patterns of vorticity in close qualitative and
quantitative resemblance with classical two-dimensional turbulence
(see e.g. Fig.~\ref{plot5}). It is also shown that such vorticity
patterns give rise to detectable current fluctuations across the
sample, well in excess of flickering noise. As a result, based on our
simulations, we conclude that the hydrodynamic picture of graphene as
a near-perfect, slow-relativistic fluid, as developed in Ref.
\cite{grapPRL}, should be liable to experimental verification.
%Should this picture be confirmed, graphene may
%open up an entirely new frontier in quantum and classical
%two-dimensional turbulence research.
\begin{figure}
  \centering
  \includegraphics[scale=0.2]{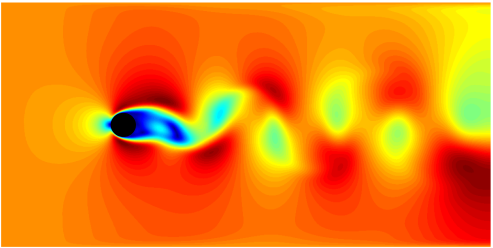}
  \caption{Vortex shedding in graphene at Reynolds number $Re=100$,
    using a grid of $1024\times 512$ cells. The color scale represents
    the absolute velocity of the fluid. The picture was taken at
    $4\times 10^6$ time steps.}\label{plot1}
\end{figure}
The equations for the Dirac electron fluid in graphene read as follows
\cite{grapPRL}: $\partial \rho_c/\partial t + \nabla \cdot \left(
  \rho_c \vec{u} \right)=0$, for charge conservation; $\partial
\epsilon/\partial t + \nabla \cdot \left[ (\epsilon+p) \vec{u} \right]
= 0$, for energy density conservation and
\begin{equation}\label{moment}
  \frac{\epsilon+p}{c^2}\left [\frac{\partial \vec{u}}{\partial t} + \left(\vec{u}\cdot \nabla \right)\vec{u} \right ]+\nabla p + \frac{\vec{u}}{c^2} \frac{\partial p}{\partial t} - \eta \nabla^2 \vec{u} = 0\quad , 
\end{equation}
for momentum conservation. Here, $c$ is the Fermi speed
($\sim10^6$m/s), $\epsilon$ the energy density, $p$ the pressure,
$\rho_c$ the charge density, and $\vec{u}$ the velocity. The shear
viscosity can be calculated by using \cite{grapPRL}
\begin{equation}\label{viscoreal}
  \eta=C_\eta \frac{N (k_B T)^2}{4 \hbar c^2 \alpha^2} \quad ,
\end{equation}
where $C_\eta \sim O(1)$ is a numerical coefficient, $T$ is the
temperature, $\alpha = e^2/\varepsilon \hbar c$ is the effective fine
structure constant, $e$ being the electric charge of the electron,
$\varepsilon$ the relative dielectric constant, and $N$ the number of
species of free massless Dirac particles. Additionally, the entropy
density can be calculated according to the Gibbs-Duhem relation
$\epsilon + p = T s$.  These equations have been derived under the
assumption $|\vec{u}| \ll c$.
%Due to the similarity between the full relativistic
%hydrodynamics and the Dirac fermions fluid equations, the extension of
%the RLB to the case of a Dirac fluid in graphene seems to be
%immediate. From now on, we will use normalize units $c=\hbar=k_B=e=1$.

The relativistic lattice Boltzmann (RLB), proposed by Mendoza et. al.
\cite{rlbPRL,rlbPRD}, is hereby adapted to reproduce, in the continuum
limit, the equations for the Dirac electron fluid described above. The
RLB model \cite{rlbPRL} was defined on a three-dimensional lattice
with nineteen discrete velocities. Since graphene is $2D$, we have
adapted the model to a two-dimensional cell with nine discrete
velocities, linking each site to its four nearest-neighbors, four
next-to-nearest neighbors (diagonal), plus a rest particle.  Two
distribution functions, $f_i$ and $g_i$, are used for the particle
number and momentum-energy, respectively. These distribution functions
evolve according to the typical Boltzmann equation in single-time
relaxation approximation \cite{BGK, rlbPRL}, $f_i(x+\delta x, t+\delta
t) - f_i(x,t) = -(f_i - f_i^{\rm eq})/\tau$ and $g_i(x+\delta x,
t+\delta t) - g_i(x,t) = -(g_i - g_i^{\rm eq})/\tau$, where $\tau$ is
the single relaxation time, and the equilibrium functions $f_i^{\rm
  eq}$ and $g_i^{\rm eq}$ are defined in Ref.~\cite{rlbPRL,rlbPRD}.
The shear viscosity, according to this model is $\eta = (\epsilon +
p)\left(\tau - \frac{1}{2} \right)c_l^2 \delta t/3 c^2$, where
$c_l=\delta x/\delta t$ is the ratio of the lattice spacing to
time-step size.

We choose the equation of state $\epsilon = 3p$, which depends on
temperature in the relativistic regime, as $\epsilon \sim (k_B T)^3/(c
\hbar)^2 = T^3$ (in normalized units $c=\hbar=k_B=e=1$)
\cite{relationeandP}. Thus, the shear viscosity $\eta$ would depend on
the third power of the temperature, leading to a different relation
than Eq.~\eqref{viscoreal}. However, in the Dirac fluid, the
relaxation time for the electrons depends on the inverse of the
temperature, $\tau_{rel}=(\hbar \alpha)^2/k_B T$ \cite{grap2PRB}, and,
therefore, introducing this dependence into the relaxation time $\tau$
of the numerical model, we obtain the correct function for the
viscosity. In numerical units ($\delta x = \delta t = c=1$), we set
the relaxation time to $\tau = \tau_0 T_0/T + 1/2$, where $T_0$ is the
initial temperature and $\tau_0$ the initial relaxation time.

The hydrodynamics equations are similar to the non-relativistic
Navier-Stokes equations with the exception of the compressibility term
$\sim \partial p/\partial t$.  This term is most likely negligible at
low frequencies, but it may become relevant at higher ones. The
Reynolds number $Re$, measuring the strength of inertial versus
dissipative terms \cite{grapPRL}, is given by $Re = (s T/c^2) (L
U_{typ}/\eta)$, where $L$ and $U_{typ}$ are the characteristic length
and flow velocity of the system, respectively. In lattice units, it
reads as
\begin{equation}\label{reynolds}
  Re = \frac{3 L U_{typ}}{ c^2 \delta t \left(\tau - \frac{1}{2} \right)}  \quad .
\end{equation}
According to classical turbulence theory, vortex shedding in graphene
is expected for Reynolds numbers well above one, typically $Re \sim 10
\div 100$. To detect signatures of pre-turbulent behavior in graphene
experiments, one can measure the fluctuations of the electric current
through the graphene sample. The current density is defined by
$\vec{j} = \rho_c \vec{u}$, and the total electric current is
calculated integrating the current density along the transverse ($y$)
coordinate. The characteristic fluctuation frequency can then be
related to the vortex shedding frequency. Macroscopic speeds $u \sim
10^5$ m/s could be achieved by the electrons in
graphene\cite{currentpossible}. The Reynolds number rewrites as: $Re =
U_{typ} L T/c^2 (\eta/s)$. According to Ref.~\cite{grapPRL}, $\eta/s$
takes values around $0.2 \hbar/k_B$, at temperature of $300$K, so that
we can write $Re = U_{typ} L / \nu_{eff}$, $\nu_{eff}= c^2 \eta/T s
\sim 0.005$ m$^2$/s being the effective kinematic viscosity.
Therefore, a sample of size $L = 5 \mu$m, within reach of current
technology, would yield $Re \sim 100$, sufficiently high to trigger
pre-turbulent phenomena, such as vortex shedding. To test the idea on
quantitative grounds, we implement a simulation on a grid with
$1024\times 512$ cells. The following initial values (numerical units)
were used: $\epsilon = 0.75$, $\rho_c=1.0$,
$\vec{u}=(u_x,0)=(0.002,0)$, and the Fermi speed $c=c_l=1.0$. The
initial value of the relaxation time was chosen $\tau_0 = 0.003$ such
that the initial shear viscosity $\eta=\frac{1}{3}\left(\tau -
  \frac{1}{2} \right) = 10^{-3}$.

A circular obstacle, with diameter $D=50$, is introduced at $(256,
256)$, modeling a $5$ micron diameter impurity in the graphene sample
(Fig.~\ref{plot1}). With this configuration, and setting $L=D$ in
Eq.~\ref{reynolds}, the Reynolds number for this system is $Re \sim
100$. We choose periodic boundary conditions at top and bottom, and
demand that the distribution functions of the boundary cells are
always equal to the equilibrium distribution functions evaluated with
the initial conditions. Free boundary conditions are imposed at the
outlet. At the left border, we set inlet conditions, where the missing
information of the distribution functions is filled by the equilibrium
distribution function corresponding to the initial conditions
\cite{succibook}. We define $\delta t = 0.05 \text{ps}$.

\begin{figure}
  \centering
  \includegraphics[scale=0.42]{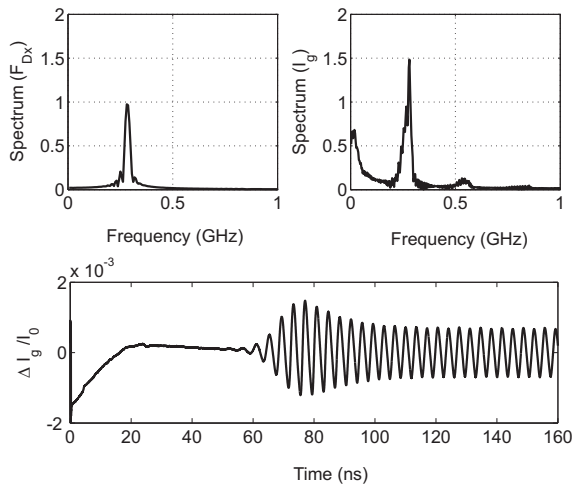}
  \caption{FFT of the electric current fluctuations $\Delta I_g$, in
    the graphene sample (top right), due to the vortex shedding as a
    function of time. Also, it is shown the FFT of the drag force
    acting on the obstacle (top left). This result refers to
    $Re=100$. At the bottom, the fluctuations in the electric current
    $I_g$ are shown as a function of time.}\label{currentFFT}
\end{figure}

The drag $F_{Dx}$ and lift $F_{Dy}$ forces acting on the obstacle are
measured, the vortex shedding frequency being computed in terms of
fluctuations of the lift forces. We compare the frequency of the
electric current fluctuations with the frequency of the drag force,
which, in general, is twice the vortex shedding frequency (see
Fig.~\ref{currentFFT}). To relate these to the vortex shedding, we use
a fast-Fourier transform (FFT). As is well visible from Fig.
\ref{currentFFT}, the current fluctuations contribute about one part
per thousand of the base signal, and, consequently, they should be
liable to experimental detection.
% ------------------------------ FIG 
% -------------------------------
In future applications, involving larger graphene samples, higher
Reynolds numbers will be attained. Consequently, it becomes of
interest to assess the role of the relativistic corrections to the
classical Navier-Stokes equations.

Comparing the dynamics of the relativistic and non-relativistic
fluids, two basic differences emerge: the relativistic correction term
$\sim \partial p /\partial t$; and the viscosity dependence with the
temperature, Eq.~\eqref{viscoreal}. In order to assess whether these
terms play an important role, we implement three simulations on a grid
of size $2048\times 1024$ cells. In the first simulation, we model the
full relativistic equations; in the second one, the relativistic
effect $\sim \partial p /\partial t$ is removed; and in the third one,
the viscosity is forced to be a constant. The same initial
configuration, as before, is used with the exception of:
$\vec{u}=(u_x,0)=(0.03,0)$ and $D=100$ (in this case, modeling an
impurity of diameter $150 \mu$m). The impurity is now centered at
$(512, 512)$. With this configuration, Eq.~\ref{reynolds} gives $Re
\sim 3000$. The simulations run up to $10^6$ time steps (with $\delta
t = 1.5 \text{ps}$).

\begin{figure}
  \centering
  \includegraphics[scale=0.35]{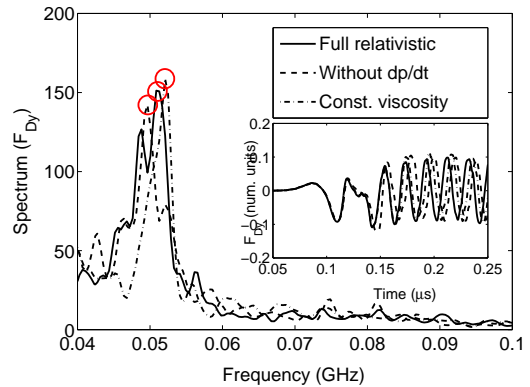}
  \caption{Frequencies of the vortex shedding at Reynolds number
    $Re=3000$, using a grid of $2048\times 1024$ cells, are
    shown. These are calculated for three different cases: the full
    relativistic, relativistic without the term $\partial p / \partial
    t$, and relativistic with constant viscosity.}\label{plot2}
\end{figure}

From Fig.~\ref{plot2}, we find that, in the case of constant
viscosity, the frequency is a bit higher than the one corresponding to
the full relativistic case. On other hand, if the term $\sim \partial
p/\partial t$ is removed from the equations, the frequency decreases.
We conclude that, in order to compare to high precision measurements of
the vortex shedding frequencies, these terms cannot be ignored.
\begin{figure}
  \centering
  \includegraphics[scale=0.35]{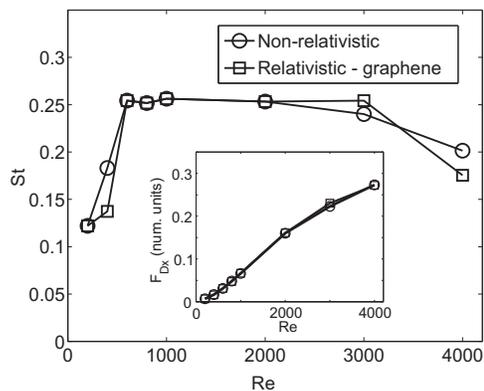}
  \caption{Strouhal number $St$ as a function of the Reynolds number
    $Re$ for both non-relativistic and relativistic fluids. In the
    inset, the mean value of the $x$-component of the drag force as a
    function of the Reynolds number is shown. The error bar is of the
    size of the symbol.}\label{plot4}
\end{figure}
\begin{figure}
  \centering
  \includegraphics[scale=0.35]{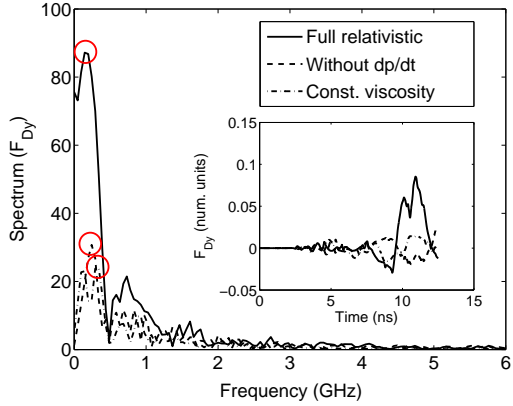}
  \caption{The same case of Fig.~\ref{plot2}, but in the case of the
    constriction at Reynolds number $Re=25$, using a grid with
    $1024\times 1024$ cells.}\label{plot2a}
\end{figure}
\begin{figure}
  \centering
  \includegraphics[scale=0.42]{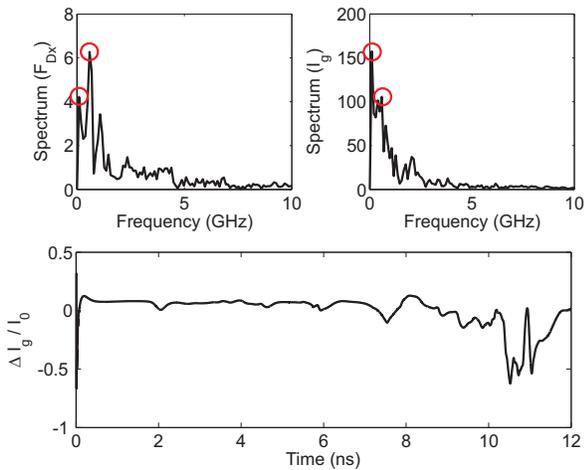}
  \caption{The same case as Fig.~\ref{currentFFT} for the constricted
    flow at $Re=25$.}\label{currentFFTa}
\end{figure}
% ----------------------------------------------------------------------------------
% -------------------------------------- FIG
\begin{figure}
  \centering
  \includegraphics[scale=0.35]{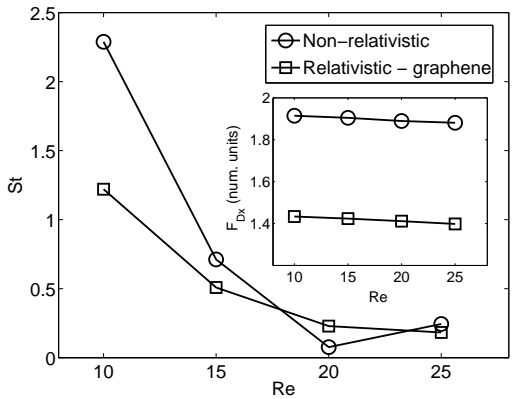}
  \caption{The same as Fig.~\ref{plot4} for the case of the
    constriction.}\label{plot4a}
\end{figure}
% ---------------------------------------------
To study the frequency of the vortex shedding, we vary the initial
velocity in order to obtain different Reynolds numbers. The Strouhal
number $St$ is defined as the dimensionless frequency of the vortex
shedding and can be calculated as $St = f_s L/U_{typ}$, where $f_s$ is
the frequency of the vortex shedding. Fig.~\ref{plot4} shows that the
relation between $St$ and $Re$ is very similar for the relativistic
and non-relativistic fluids, with a fast growth of $St$ in the range
$200<Re<1000$, followed by a flat-top at $St \sim 0.25$
\cite{St1,St2,St3} for $Re>1000$. From the Strouhal number, we can
obtain the frequencies of the vortex shedding, as $f_s = 0.2
U_{typ}/L$. The frequency of the drag force is twice that of vortex
shedding, namely $f_{Ds} = 0.4 U_{typ}/L$.  As a result, once the
Reynolds number is known, one can compute the frequency of the drag
force, the Strouhal number, and then compare with the FFT of the
electric current measurement in the sample. The mean value of the drag
force $\overline{F}_{Dx}$, reported in the inset of Fig.~\ref{plot4}
as a function of the Reynolds number, shows a monotonic dependence in
the range of $Re$ explored here.

Another kind of set-up to detect pre-turbulence in graphene
experiments, with the possibility of being implemented nowadays,
consists of building a constriction, where the Dirac fluid can develop
vorticity patterns as it crosses through. Fig.~\ref{plot5} shows the
vorticity at $Re = 25$, where the characteristic length $L = 50$ cells
has been chosen as the distance between the tips. In this case, the
initial velocity is taken $\vec{u} = (u_x,0) = (0.0005,0)$, in lattice
units, and the simulation is performed using a grid of $1024\times
1024$ cells. We simulate two systems, one with the full relativistic
equations and the other one by just removing the relativistic term
$\partial p/ \partial t$. From the simulations (see Fig.~\ref{plot5}),
we conclude that the relativistic contribution affects the time to the
onset of instability, and, from Fig.~\ref{plot2a}, we can appreciate
that, as for the circular impurity, the frequency of the vortices
presents a shift due to the relativistic corrections. However, both
constant viscosity and removal of the relativistic correction,
contribute to an increase of the frequency of the fluctuations.
Fig.~\ref{currentFFTa} shows how such fluctuations can be measured,
and the characteristic frequencies (see red circles in
Fig.~\ref{currentFFTa}) related with the drag force acting on the
constriction. Note that, in order to achieve $Re = 25$, at a speed of
$0.1 c$, the distance between tips is about $1.25 \mu$m.

As for the case of the circular impurity, we can find the
characteristic relation between the Strouhal number and the Reynolds
number for this geometrical set-up (see Fig.~\ref{plot4a}). From the
inset of Fig.~\ref{plot4a}, we observe that the drag force decreases
slightly, as the Reynolds number is increased, and exhibits a
noticeable difference between the non-relativistic and relativistic
cases.

Summarizing, we have shown that, in the range of $Re \sim 10^2$,
vorticity patterns can be indirectly observed by measuring the
electric current fluctuations in the graphene sample. However, using a
different geometry, like a constriction, signatures of pre-turbulence
can be detected already at Reynolds numbers as small as $Re \sim 25$.
We have also compared the effects of relativistic corrections, such as
dynamic compressibility and the dependency of the viscosity on the
temperature, on the dynamics of the system. In these cases, the
temperature dependency of the viscosity and the term $\partial
p/\partial t$ produce a shift in the frequencies of the vortex
shedding and, therefore, in the electric current fluctuations.
Additionally, the relativistic correction term, $\sim
\partial p/ \partial t$, is found to delay the instability process in
the case of the constricted flow. For future applications, most likely
accessing higher Reynolds numbers, the frequency of the vortex
shedding can be calculated using the Strouhal number, thereby
permitting to distinguish current fluctuations induced by
pre-turbulent phenomena from those resulting from other physical
effects.

We thank K. Ensslin for enlightening discussions.

\bibliography{report}

\end{document}